\documentclass[aps,prl,twocolumn]{revtex4-2}

\usepackage{graphicx}
\usepackage{amsmath}
\usepackage{bm}

\begin{document}

\title{Flexoelectricity versus Electrostatics in Polar Nematic Liquid Crystals}
\author{Lincoln Paik}
\author{Jonathan V. Selinger}
\affiliation{Department of Physics, Advanced Materials and Liquid Crystal Institute, Kent State University, Kent, Ohio 44242, USA}
\date{\today}
\begin{abstract}
Polar nematic liquid crystals have two special features, compared with conventional nematic liquid crystals.  First, because of flexoelectricity, the combination of polar order and splay reduces the free energy.  Second, because of electrostatics, any splay generates a bound charge density, which increases the free energy.  To assess the competition between these two effects, we develop a theory that combines flexoelectricity and electrostatics.  The theory predicts a phase diagram that includes ferroelectric, antiferroelectric, and conventional nematic phases.
\end{abstract}

\maketitle

One important development in recent liquid-crystal research is the experimental discovery of the ferroelectric nematic phase~\cite{Chen2020,Lavrentovich2020,Sebastian2022}.  The conventional nematic phase has orientational order with two-fold symmetry, without polarity, but the new phase has a large spontaneous electrostatic polarization.  This polarization has profound effects on the response to applied electric fields, and on the director configurations even in the absence of applied fields~\cite{Basnet2022,Zavvou2022,Yang2022,Kumari2023,Perera2023,Yang2024,Kumari2024}.  Along with the conventional and ferroelectric nematic phases, the same materials often show an intermediate antiferroelectric phase with alternating polarization, sometimes called a splay nematic or \hbox{smectic-$Z_A$} phase~\cite{Mertelj2018,Mandle2019,Connor2020,Chen2023}.  This intermediate phase is closely associated with the ferroelectric phase, indicating that antiferroelectricity and ferroelectricity are two related manifestations of polar order in nematic liquid crystals.

From a theoretical perspective, polar nematic liquid crystals have two special features---flexoelectricity and electrostatics---which both differ from conventional nematic liquid crystals.  As discussed below, these two features have conflicting effects on the director configuration.  In a polar phase, flexoelectricity favors splay, but electrostatics creates a strong energy penalty for splay.  Hence, an important issue for this field is how to assess the competition between these two effects.

The purpose of this paper is to combine flexoelectricity and electrostatics into a single theory for polar nematic liquid crystals.  We first construct a free energy that includes both of these effects.  Using this free energy, we revisit the classic calculation of Khachaturyan for spontaneous cholesteric twist arising from electrostatic interaction~\cite{Khachaturyan1975}.  We then consider how previous theoretical research on the splay nematic phase~\cite{Mertelj2018,Copic2020,Rosseto2020} is modified by the combination of flexoelectricity and electrostatics.  This work leads to a phase diagram with ferroelectric, antiferroelectric, and conventional nematic phases.

The flexoelectric effect is a coupling between splay deformations and polar order parallel to the director (or between bend deformations and polar order perpendicular to the director)~\cite{Meyer1969}.  In this effect, if one applies a splay to a conventional nematic liquid crystal, the material develops polar order, which may be detected as a net electrostatic polarization.  Conversely, if one applies an electric field to the liquid crystal, the material develops splay.  Although the effect can be detected as an electric polarization, it is not caused by electrostatics.  Rather, it is an elastic coupling, which is required by symmetry.  This coupling \emph{reduces} the effective splay elastic constant $K_{11}$~\cite{Meyer1976,Selinger2022}.

Flexoelectricity is particularly important for polar nematic liquid crystals, in which the polar order occurs spontaneously.  Because of this spontaneous polar order, the optimal local state of the director field has nonzero splay.  In that case, the liquid crystal experiences geometric frustration, in that it cannot fill up space with the optimal local structure~\cite{Selinger2022}.  As a result, the material can break into domains of splay in alternating directions, separated by domain walls.  The domains may be arranged in a one-dimensional (1D) array~\cite{Mertelj2018,Copic2020}, or in a 2D square lattice like a checkerboard~\cite{Rosseto2020}.

In addition to flexoelectricity, electrostatic interactions provide another important contribution to the free energy of polar nematic liquid crystals.  A polar phase has an electrostatic polarization density $\bm{P}(\bm{r})$, which is at least approximately aligned with the director field $\hat{\bm{n}}(\bm{r})$.  Inside the material, any splay of the polarization gives a bound electric charge density $\rho_\text{bound}(\bm{r})=-\bm{\nabla}\cdot\bm{P}$.  Furthermore, at the surface of the material, there is a surface electric charge density $\rho_\text{surface}(\bm{r})=(\hat{\bm{N}}\cdot\bm{P})\delta(\hat{\bm{N}}\cdot\bm{r})$, where $\hat{\bm{N}}$ is the local unit vector normal to the surface.  This electric charge density interacts through a Coulomb potential, possibly screened by counterions, and generates a high electrostatic energy.  For that reason, there is a severe energy penalty for any splay of $\bm{P}$, and hence for any splay of $\hat{\bm{n}}$.  This electrostatic energy \emph{increases} the effective splay elastic constant $K_{11}$~\cite{Okano1986,Basnet2022,Zavvou2022}.

To understand polar nematic liquid crystals with both flexoelectricity and electrostatics, we must develop a theory that includes both of these effects.  We use the free energy $F=F_\text{Frank}+F_\text{Landau}+F_\text{flexo}+F_\text{elec}$.  The first part is the Frank free energy,
\begin{align}
F_\text{Frank}&=\int d^3 r \biggl[\frac{1}{2}K_{11}(\bm{\nabla}\cdot\hat{\bm{n}})^2
+\frac{1}{2}K_{22}(\hat{\bm{n}}\cdot\bm{\nabla}\times\hat{\bm{n}})^2\nonumber\\
&\qquad\qquad+\frac{1}{2}K_{33}(\hat{\bm{n}}\times(\bm{\nabla}\times\hat{\bm{n}}))^2\biggr].
\end{align}
The second part is the Landau free energy as a power series in the polarization,
\begin{align}
F_\text{Landau}&=\int d^3 r \biggl[\frac{1}{2}\mu_\parallel(\bm{P}\cdot\hat{\bm{n}})^2
+\frac{1}{2}\mu_\perp|\bm{P}\times\hat{\bm{n}}|^2\nonumber\\
&\qquad\qquad+\frac{1}{4}\nu|\bm{P}|^4
+\frac{1}{2}\kappa|\bm{\nabla}\bm{P}|^2\biggr].
\end{align}
The quadratic term has previously been written as just $\frac{1}{2}\mu|\bm{P}|^2$~\cite{Rosseto2020,Selinger2022}, but here we explicitly consider different coefficients for components of $\bm{P}$ parallel and perpendicular to the local director $\hat{\bm{n}}$.  We anticipate that $\mu_\parallel$ will pass through zero as a function of temperature, leading to polar order parallel to $\hat{\bm{n}}$.  By contrast, $\mu_\perp$ will be a fixed energy penalty---large, positive, and roughly constant with respect to temperature.  The third part of the free energy is the flexoelectric coupling between polarization and splay,
\begin{equation}
F_\text{flexo}=\int d^3 r \left[-\lambda(\bm{P}\cdot\hat{\bm{n}})(\bm{\nabla}\cdot\hat{\bm{n}})\right],
\end{equation}
as has been discussed in previous work~\cite{Rosseto2020,Selinger2022}.  The fourth part is the screened electrostatic interaction,
\begin{equation}
F_\text{elec}=\frac{1}{2}\int d^3 r_1 d^3 r_2 \frac{\rho(\bm{r}_1)\rho(\bm{r}_2) e^{-|\bm{r}_{12}|/\Lambda}}{4\pi\epsilon |\bm{r}_{12}|},
\end{equation}
which couples the electric charge density $\rho=\rho_\text{bound}+\rho_\text{surface}$ at positions $\bm{r}_1$ and $\bm{r}_2$, with $\bm{r}_{12}=\bm{r}_2-\bm{r}_1$.  The parameter $\Lambda$ is the Debye screening length, with $\Lambda\to\infty$ corresponding to an unscreened Coulomb interaction, and $\epsilon$ is the dielectric constant.  The prefactor of $\frac{1}{2}$ avoids double-counting.

\begin{figure}
(a)\includegraphics[width=0.44\columnwidth]{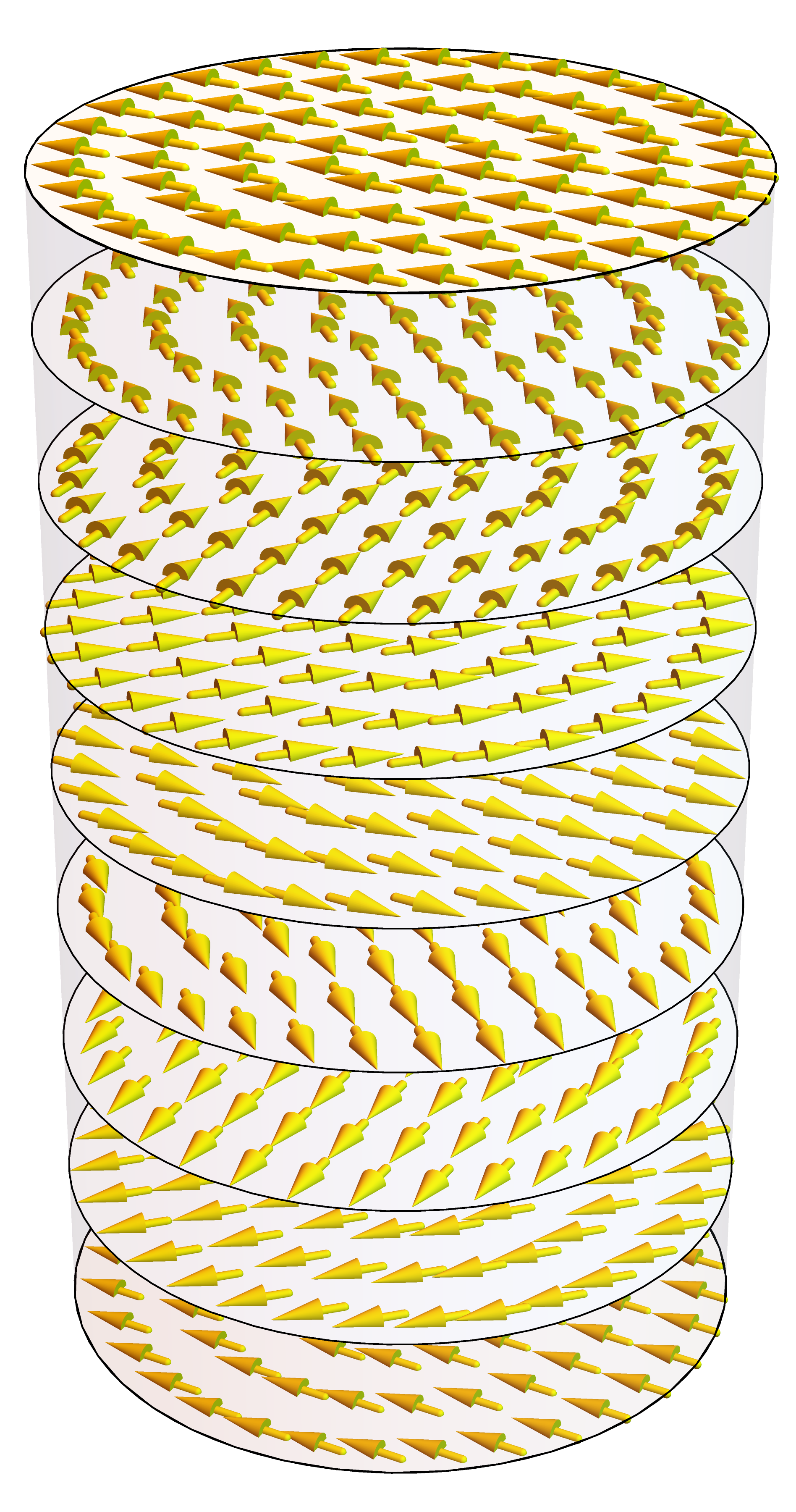}
(b)\includegraphics[width=0.44\columnwidth]{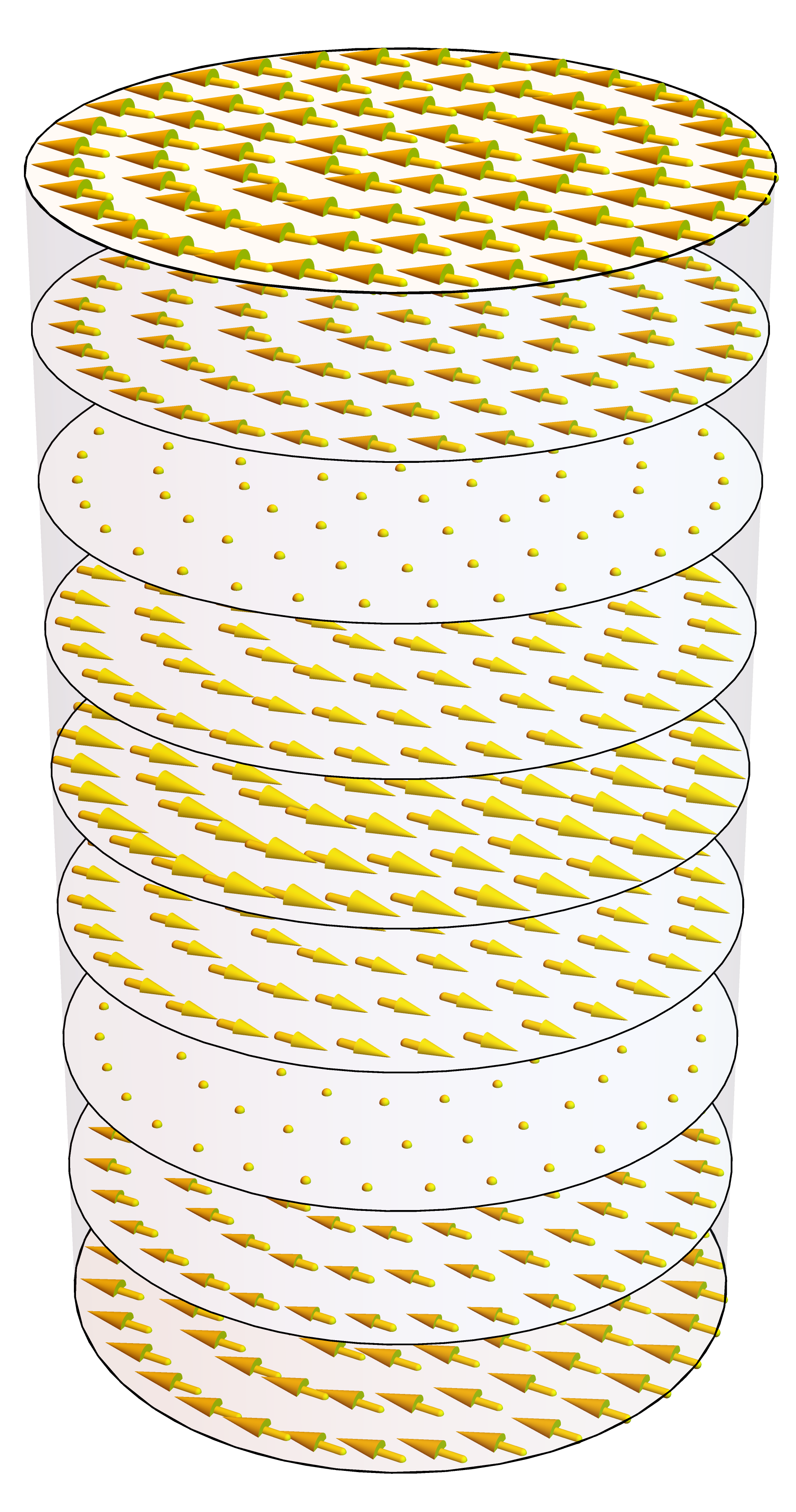}
\caption{Configurations of polar order studied in this paper.  (a)~Spontaneous cholesteric.  (b)~Splay nematic.}
\label{geometry}
\end{figure}

As a preliminary application of this free energy, we consider the possibility of spontaneous cholesteric twist.  In 1975, Khachaturyan predicted that a polar nematic liquid crystal with free boundaries would be unstable to the formation of a cholesteric helix, randomly right- or left-handed, in order to reduce the electrostatic interaction energy~\cite{Khachaturyan1975}.  Recently, Kumari et al.\ reported experimental observations of this type of helix~\cite{Kumari2024}.  They also pointed out a typographical error in Khachaturyan's paper, with the wrong dimensions in the prediction for the cholesteric pitch.

To explore this concept, we consider a polar nematic liquid crystal in the cylindrical geometry shown in Fig.~\ref{geometry}(a), with length $L\to\infty$, and large but finite radius $R$.  As an ansatz, we suppose the director has the twisted configuration $\hat{\bm{n}}=(\cos q z,\sin q z,0)$, and the polarization is $\bm{P}=P_0\hat{\bm{n}}$, for fixed magnitude $P_0$.  By putting this ansatz into the free energy functional, and assuming $q R \gg 1$, we obtain the free energy per volume
\begin{equation}
\frac{F}{\pi R^2 L}=\frac{1}{2}\mu_\parallel P_0^2 +\frac{1}{4}\nu P_0^4 +\frac{1}{2}K_{22}^\text{eff}q^2
+\frac{P_0^2}{4\epsilon R\sqrt{q^2+\Lambda^{-2}}},
\end{equation}
with $K_{22}^\text{eff}=K_{22}+\kappa P_0^2$.  The flexoelectric term is zero, because the ansatz does not have any splay.  Note that the electrostatic term depends on the radius $R$.  This dependence on system size does not usually occur in elasticity theory, but it is normal in electrostatics, because of the long-range interactions.

We can now minimize the free energy over the helical twist $q$.  For this minimization, we must distinguish between two regimes.  In the regime of weak electrostatics, with $\Lambda<\Lambda_c=(4K_{22}^\text{eff}\epsilon R/P_0^2)^{1/3}$, the minimum occurs at $q=0$, or pitch
$2\pi/q\to\infty$.  Hence, the system forms a uniform ferroelectric nematic phase.  By contrast, in the regime of strong electrostatics, with $\Lambda>\Lambda_c=(4K_{22}^\text{eff}\epsilon R/P_0^2)^{1/3}$, the minimum occurs at $q=(\Lambda_c^{-2}-\Lambda^{-2})^{1/2}$, or at pitch $2\pi/q=2\pi(\Lambda_c^{-2}-\Lambda^{-2})^{-1/2}$.  This result corrects the typographical error in Ref.~\cite{Khachaturyan1975}.  A similar calculation has recently been done by M.~O.\ Lavrentovich~\cite{Lavrentovich2024}.

Next, we use the same free energy to model a splay nematic phase.  In particular, we consider a 1D splay nematic phase, not the 2D splay nematic phase proposed in Ref.~\cite{Rosseto2020}.  In this structure, the polar order alternates in sign along a single axis.  We can describe this alternation as a sinusoidal modulation $\bm{P}(z)=(P_0\cos q z,0,0)$, as illustrated in Fig.~\ref{geometry}(b).  For the corresponding director modulation, we assume the structure with alternating splay $\hat{\bm{n}}(z)=(\cos\theta(z),0,\sin\theta(z))$, with $\theta(z)=\theta_0\sin q z$.  This form is similar but not identical to the assumption in previous studies of the 1D splay nematic phase~\cite{Mertelj2018,Copic2020,Rosseto2020}.  The slight difference is that the polarization is not strictly parallel to the nematic director.  In the current theory, we make a simplifying assumption that $\bm{P}$ is in the $x$-direction, but we allow $\hat{\bm{n}}$ to vary in the $(x,z)$-plane.  The $\mu_\perp$ term in the free energy aligns $\bm{P}$ and $\hat{\bm{n}}$, and keeps the deviations from growing too large.

We put the assumptions for $\bm{P}$ and $\hat{\bm{n}}$ into the free energy, and average over $z$, to obtain the average free energy per volume
\begin{align}
\frac{F}{\pi R^2 L}&=\frac{[K_{11}q^2\theta_0^2+\mu_\parallel P_0^2-2\lambda q P_0 \theta_0]
[\theta_0+J_1(2\theta_0)]}{8\theta_0}\nonumber\\
&\quad+\frac{[K_{33}q^2\theta_0^2+\mu_\perp P_0^2][\theta_0-J_1(2\theta_0)]}{8\theta_0}\nonumber\\
&\quad+\frac{3\nu P_0^4}{32}+\frac{\kappa q^2 P_0^2}{4}
+\frac{P_0^2}{8\epsilon R\sqrt{q^2+\Lambda^{-2}}},
\end{align}
using the Bessel function $J_1(2\theta_0)$.  This average free energy is expressed in terms of three variational parameters:  the director amplitude $\theta_0$, polarization amplitude $P_0$, and wavevector $q$.  To analyze it, we first minimize over the director amplitude, and obtain $\theta_0$ as a power series in $P_0$,
\begin{equation}
\theta_0=\frac{\lambda P_0}{K_{11}q}
-\frac{(A+\lambda^2 K_{33})\lambda P_0^3}{4K_{11}^4 q^3}+O(P_0^5).
\end{equation}
By putting that result back into the free energy, we obtain an effective free energy in terms of $P_0$ and $q$ only,
\begin{align}
\frac{F}{\pi R^2 L}&=\frac{1}{4}\left[\mu_\parallel-\frac{\lambda^2}{K_{11}}+\kappa q^2
+\frac{1}{2\epsilon R\sqrt{q^2+\Lambda^{-2}}}\right]P_0^2\nonumber\\
&\quad+\left[\frac{3\nu}{32}+\frac{A\lambda^2}{16K_{11}^4 q^2}\right]P_0^4
+O(P_0^6).
\label{Feff}
\end{align}
In both of these expressions, we define
\begin{equation}
A=\lambda^2(K_{11}+K_{33})+K_{11}^2(\mu_\perp-\mu_\parallel).
\end{equation}

To interpret this free energy, consider the coefficient of the quadratic term in the power series.  As usual in the theory of phase transitions, a second-order transition from $P_0=0$ to $P_0\neq0$ occurs when the quadratic coefficient changes sign.  In the quadratic coefficient, the term $\mu_\parallel$ is a temperature-like variable.  At high temperature, $\mu_\parallel$ is positive because of the entropic penalty for polar order.  At low temperature, $\mu_\parallel$ is negative because of the local energetic benefit of polar order.  Flexoelectricity provides a negative correction of $\lambda^2/K_{11}$, which shifts the transition to higher values of $\mu_\parallel$, and hence increases the temperature range of the polar order.  Electrostatics provides a positive correction of $1/(2\epsilon R\sqrt{q^2+\Lambda^{-2}})$, which shifts the transition to lower values of $\mu_\parallel$, and hence reduces the temperature range of polar order.

In the quadratic coefficient, the elastic and electrostatic terms both depend on wavevector $q$.  The elastic term $\kappa q^2$ favors small $q$, and the electrostatic term favors large $q$.  By minimizing their sum over $q$, we can see that there are distinct regimes of weak and strong electrostatics.  In the regime of weak electrostatics, with Debye screening length $\Lambda<\Lambda_c=(4\kappa\epsilon R)^{1/3}$, the quadratic term first passes through zero at wavevector $q_{c1}=0$.  Hence, the nonpolar state becomes unstable to the formation of polar order with infinite wavelength, at the critical value of $\mu_\parallel$ given by
\begin{equation}
\mu_{c1}=\frac{\lambda^2}{K_{11}}-\frac{\Lambda}{2\epsilon R}.
\end{equation}
By contrast, in the regime of strong electrostatics, with $\Lambda>\Lambda_c=(4\kappa\epsilon R)^{1/3}$, the quadratic term first passes through zero at wavevector
\begin{equation}
q_{c2}=(\Lambda_c^{-2}-\Lambda^{-2})^{1/2}.
\end{equation}
In this case, the nonpolar state becomes unstable at a finite wavelength, at the critical value of $\mu_\parallel$ given by
\begin{equation}
\mu_{c2}=\frac{\lambda^2}{K_{11}}+\frac{\kappa}{\Lambda^2}-\frac{3\kappa^{1/3}}{(4\epsilon R)^{2/3}}.
\end{equation}

\begin{figure}
\includegraphics[width=\columnwidth]{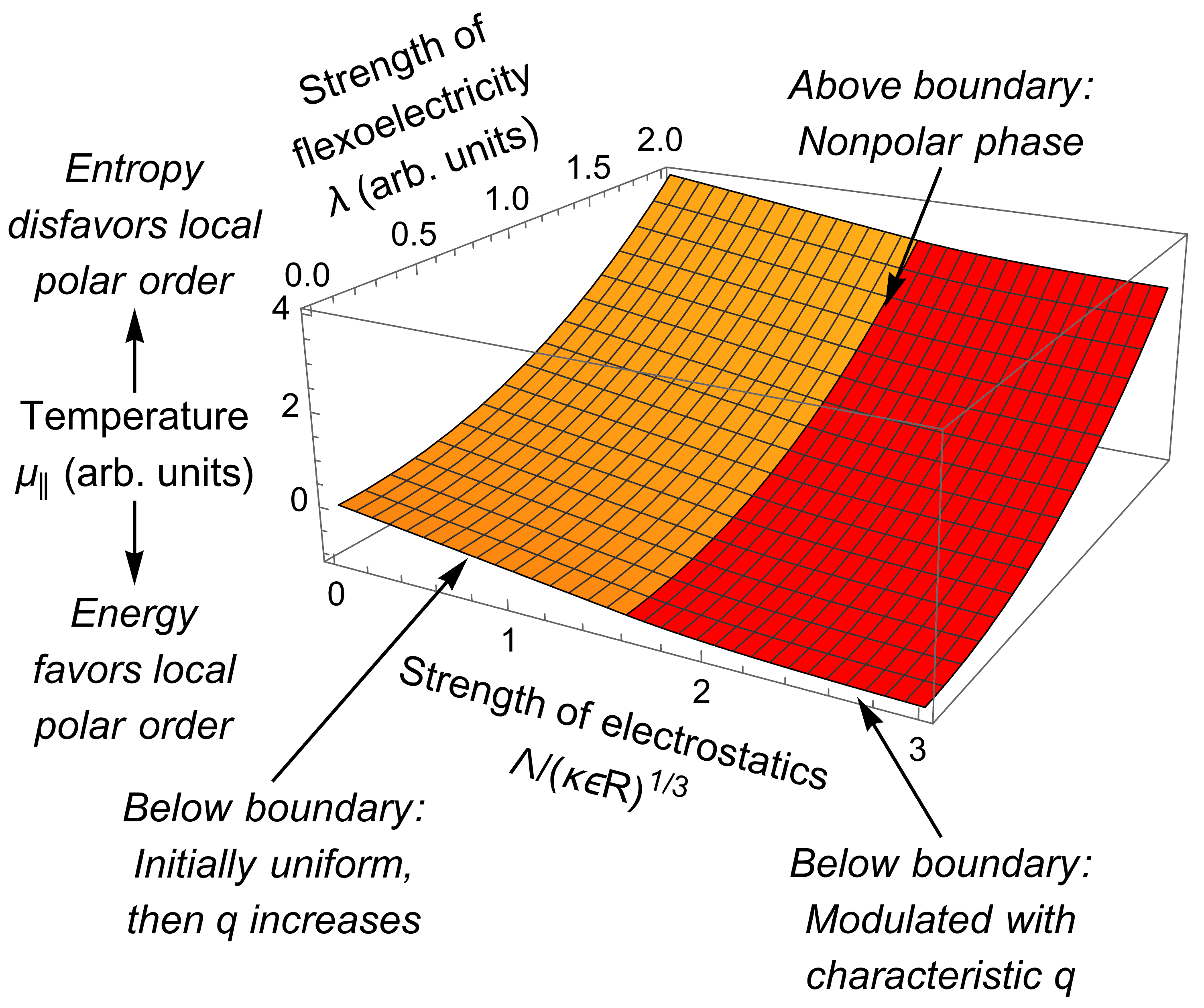}
\caption{Phase diagram for the model discussed here, in arbitrary units with $K=\kappa=\epsilon=R=1$.  Above the surface, the system is in the conventional nonpolar nematic phase.  Below the surface, the system is in a polar phase.  The orange part of the surface is the regime of weak electrostatics, in which the polar order begins at $q=0$.  The red part of the surface is the regime of strong electrostatics, in which the polar order begins with a characteristic finite $q$.}
\label{phasediagram}
\end{figure}

These results are summarized in the 3D phase diagram of Fig.~\ref{phasediagram}.  Here, one horizontal axis is the scaled Debye screening length $\Lambda/(\kappa\epsilon R)^{1/3}$, which can be interpreted as the strength of the electrostatic interaction.  The other horizontal axis is the coefficient $\lambda$, which can be interpreted as the strength of the flexoelectric coupling.  The vertical axis is $\mu_\parallel$, which is a temperature-like variable.  At high $\mu_\parallel$, local polar order is disfavored by entropy.  At low $\mu_\parallel$, local polar order is favored by energy.  The orange and red surface represents the boundary between a conventional nonpolar nematic phase (above the surface) and a polar nematic phase (below the surface).  This phase boundary goes to lower temperature with increasing strength of electrostatics, and it goes to higher temperature with increasing strength of flexoelectricity.  The orange part of the surface is the regime of weak electrostatics.  Below this transition, the polar nematic phase is initially uniform with wavevector $q_{c1}=0$.  The red part of the surface is the regime of strong electrostatics.  Below that transition, the polar nematic phase begins as an antiferroelectric, modulated structure with the characteristic wavevector $q_{c2}$.

To analyze the mean-field critical behavior below the transition, we can minimize the free energy of Eq.~(\ref{Feff}).  First, consider the regime of weak electrostatics $\Lambda<\Lambda_c=(4\kappa\epsilon R)^{1/3}$, below the orange surface.  We first minimize the free energy over wavevector $q$, and obtain $q$ as a power series in $P_0$, with the form 
\begin{equation}
q=\frac{(R\epsilon A\lambda^2 P_0^2)^{1/4}}{K_{11}(\Lambda_c^3-\Lambda^3)^{1/4}}
-\frac{3\Lambda^5(R\epsilon A\lambda^2 P_0^2)^{3/4}}{8K_{11}^3(\Lambda_c^3-\Lambda^3)^{7/4}}
+O(P_0^{5/2}).
\end{equation}
We then put that series back into the free energy, to obtain the effective free energy in terms of $P_0$ alone,
\begin{align}
\frac{F}{\pi R^2 L}&=\frac{1}{4}(\mu_\parallel-\mu_{c1})P_0^2
+\frac{[A\lambda^2(\Lambda_c^3-\Lambda^3)]^{1/2}}{8K_{11}^2 (R\epsilon)^{1/2}}|P_0|^3\nonumber\\
&\quad+\left[\frac{3\nu}{32}+\frac{3A\lambda^2\Lambda^5}{64K_{11}^4(\Lambda_c^3-\Lambda^3)}\right]P_0^4
+O(P_0^5).
\end{align}
This form of the free energy is unusual in the theory of phase transitions, because the $|P_0|^3$ term is nonanalytic at $P_0=0$.  That nonanalytic term is not permitted in the input to Landau theory, but it can occur in the output from Landau theory, after eliminating other variables.  Here, it occurs after minimization over $\theta_0$ and $q$.  The cubic term does not lead to a first-order transition: because it involves the absolute value of $P_0$, it does not induce a stable or metastable minimum of the free energy at large $P_0$.  Rather, it leads to the mean-field scaling behavior
\begin{equation}
P_0\sim(\mu_{c1}-\mu)^1,\
\theta_0\sim(\mu_{c1}-\mu)^{1/2},\
q\sim(\mu_{c1}-\mu)^{1/2}.
\end{equation}
This behavior is the same scaling that has been predicted in previous studies of the splay nematic phase~\cite{Mertelj2018,Copic2020,Rosseto2020}.  The prediction for $q$ implies that the polar phase is initially uniform (ferroelectric nematic) right at the transition, and then it becomes modulated with increasing $q$ as the temperature decreases below the transition.

Next, consider the regime of strong electrostatics $\Lambda>\Lambda_c=(4\kappa\epsilon R)^{1/3}$, below the red surface.  We minimize the free energy over $q$, as a power series in $P_0$, and obtain
\begin{equation}
q=q_{c2}+\frac{A\lambda^2 P_0^2}{12\kappa K_{11}^4\Lambda_c^2(\Lambda_c^{-2}-\Lambda^{-2})^{5/2}}+O(P_0^4).
\end{equation}
We put that result back into the free energy, to obtain the effective free energy in terms of $P_0$ alone,
\begin{align}
\frac{F}{\pi R^2 L}&=\frac{1}{4}(\mu_\parallel-\mu_{c2})P_0^2\\
&\quad+\left[\frac{3\nu}{32}+\frac{A\lambda^2}{16K_{11}^4(\Lambda_c^{-2}-\Lambda^{-2})}\right]P_0^4
+O(P_0^6).\nonumber
\end{align}
This form of the free energy is a more conventional expression in the theory of phase transitions, and it is analytic in $P_0$.  It leads to the mean-field scaling behavior
\begin{equation}
P_0\sim(\mu_{c2}-\mu)^{1/2},\
\theta_0\sim(\mu_{c2}-\mu)^{1/2},\
q\sim q_{c2}.
\end{equation}
In this case, the polar phase is modulated (antiferroelectric) right at the transition, and the modulation wavelength is approximately constant with respect to temperature below the transition.

One limitation of this theory is that it assumes that the director and polar order are modulated in simple sine and cosine waves.  From previous research on modeling the splay nematic phase without electrostatics~\cite{Rosseto2020}, we know that this assumption is only valid near the critical temperature.  Farther below the critical temperature, the modulation changes:  Instead of a sinusoidal form, it becomes a series of well-defined domains (with approximately constant polar order and nonzero splay) separated by sharp domain walls (in which the polar order passes through zero and changes sign).  As the temperature decreases, the size of the domains increases, and eventually the system crosses over into a uniform polar phase.  This trend should occur also in the presence of electrostatic interactions, leading to a uniform polar (ferroelectric nematic) phase well below the boundary shown in Fig.~\ref{phasediagram}.

In conclusion, this paper has combined the two special features of polar nematic liquid crystals---flexoelectricity and electrostatics---into a single theory.  This combination gives a phase diagram for nonpolar (conventional nematic), modulated polar (antiferroelectric), and uniform polar (ferroelectric nematic) states.  The theory shows that flexoelectricity and electrostatics often have opposite effects from each other.  Flexoelectricity tends to favor splay and polar order, so that the effective elastic constant $K_{11}$ is reduced and the temperature range of a polar phase is increased.  Electostatics tends to disfavor splay and polar order, so that the effective $K_{11}$ is increased and the temperature range of a polar phase is reduced.  These general considerations may help to interpret experiments on polar nematic liquid crystals.

We thank M.~O. Lavrentovich, O.~D. Lavrentovich, and L. Radzihovsky for helpful discussions.  This work was supported by National Science Foundation Grant DMR-1409658.

\bibliography{version2}

\end{document}